# Stellar feedback as the origin of an extended molecular outflow in a starburst galaxy

J. E. Geach[1], R. C. Hickox[2], A. M. Diamond-Stanic[3], M. Krips[4], G. H. Rudnick[5,6], C. A. Tremonti[3], P. H. Sell[7], A. L. Coil[8], J. Moustakas[9]

[1] Centre for Astrophysics Research, University of Hertfordshire, College Lane, Hatfield, Hertfordshire, AL10 9AB, UK
[2] Department of Physics and Astronomy, Dartmouth College, Hanover, NH 03755, USA
[3] Department of Astronomy, University of Wisconsin-Madison, Madison, WI 53706, USA
[4] Institut de Radioastronomie Millimétrique, 300 rue de la Piscine F-38406 Saint Martin d'Hères, France
[5] Department of Physics and Astronomy, University of Kansas, Lawrence, KS 66045, USA
[6] Max Planck Institute for Astronomy, Königstuhl 17, D-69117, Germany
[7] Department of Physics, Texas Tech University, Lubbock, TX 79409-1051, USA
[8] Center for Astrophysics and Space Sciences, University of California, San Diego, La Jolla, CA 92093, USA
[9] Department of Physics and Astronomy, Siena College, 515 Loudon Road, Loudonville, NY 12211, USA

**Recent observations have revealed that starburst galaxies can drive molecular gas outflows through stellar radiation pressure[1,2]. Molecular gas is the phase of the interstellar medium (ISM) from which stars form, so these outflows curtail stellar mass growth in galaxies. Previously known outflows, however, involve small fractions of the total molecular gas content and are restricted to sub-kiloparsec scales[1,2]. It is also apparent that input from active galactic nuclei is in at least some cases dynamically important[2,3], so pure stellar feedback has been considered incapable of aggressively terminating star formation on galactic scales. Extraplanar molecular gas has been detected in the archetype starburst galaxy M82[4,5], but so far there has been no evidence that starbursts can propel significant quantities of cold molecular gas to the same galactocentric radius (~10 kpc) as the warmer gas traced by metal absorbers[6,7]. Here we report observations of molecular gas in a compact (effective radius 100 pc) massive starburst galaxy at z=0.7, which is known to drive a fast outflow of ionized gas[8]. We find that a total of 35 per cent of the total molecular gas is spatially extended on a scale of approximately 10 kpc, and one third of this has a velocity of up to 1000 km s$^{-1}$. The kinetic energy associated with this high-velocity component is consistent with the momentum flux available from stellar radiation pressure[9–12]. This result demonstrates that nuclear bursts of star formation are capable of ejecting large amounts of cold gas from the central regions of galaxies, thereby strongly affecting their evolution[13,14].**

SDSS J0905+57 ($z = 0.712$) is a compact starburst galaxy with emission line properties consistent with a star-forming galaxy and no observational evidence of strong hot dust continuum in the mid-infrared, indicating no significant black hole accretion activity[8,15]. The galaxy is driving a wind with one of the highest velocities known for any star-forming galaxy, with the interstellar absorption lines of Ca II, Fe II and Mg II blueshifted by 2500 km s$^{-1}$ with respect to the Balmer stellar absorption lines. The total infrared luminosity is $L_{\rm IR} \approx 1.2 \times 10^{39}$ W corresponding to a star formation rate of $260\,M_\odot\,{\rm yr}^{-1}$. Hubble Space Telescope observations reveal that SDSS J0905+57 is extremely compact in the rest-frame $V$-band (475 nm), with an effective radius of $r_{\rm e} = 94$ pc (comparable to the size of 30 Doradus). This implies a star formation rate density $\Sigma_{\rm SFR} \approx 4700\,M_\odot\,{\rm yr}^{-1}\,{\rm kpc}^{-2}$. The compact nature of the galaxy and the high density of central star formation suggest that SDSS J0905+57 is likely to be at the final stage of a major merger[15] and is the progenitor of an elliptical galaxy.

We observed SDSS J0905+57 with the Institut de Radioastronomie Millimétrique Plateau de Bure Interferometer in the 2 mm band with receivers tuned to the frequency of the redshifted CO(2–1) emission line at $z = 0.712$ (134 GHz). At temperatures of the order 10 K, the carbon monoxide $J = 2 \rightarrow 1$ rotational transition is excited at a critical density of $n_{\rm H} \sim 10^4\,{\rm cm}^{-3}$ and is a good tracer of the bulk of the cold molecular gas reservoir. The spectrum (Figure 1) reveals a detection of the CO(2–1) emission line at $\nu_{\rm obs} = 134.666$ GHz, corresponding to $z_{\rm CO} = 0.712$, consistent with the stellar redshift, with the full width at half maximum approximately $200\,\text{km}\,\text{s}^{-1}$. We refer to this as the 'core' line. The spectrum also reveals CO emission in a broad wing extending up to $1000\,\text{km}\,\text{s}^{-1}$ from the core line. When averaged over $\Delta V = 200\text{--}1000\,\text{km}\,\text{s}^{-1}$, the emission is significant and peaks $(1.2 \pm 0.3)''$ or $(8 \pm 2)\,\text{kpc}$ from the core line with a flux density of $S = 0.43 \pm 0.09\,\text{mJy}$ (Figure 2, 3). We interpret these observations as evidence that molecular gas is being driven out of the galaxy through stellar feedback processes.

The core CO line emission is also marginally resolved beyond the $3''$ beam when averaged over the full core line width (full width at zero intensity, $\Delta V = 400\,\text{km}\,\text{s}^{-1}$, Figure 2). To confirm this, and to measure the size of the extended CO emission, we examine the *uv* plane visibilities to evaluate the average signal amplitude as a function of baseline separation (Methods). The data are inconsistent with a flat profile that would indicate an unresolved source but are better fit by a combination of a point source and circular gaussian profile with a half power radius of $1.6^{+0.8}_{-0.4}$ arcseconds. A point source-only model can be ruled out at the $4.7\sigma$ level. The angular size of this extended component corresponds to a radius of $12^{+6}_{-3}\,\text{kpc}$ in physical projection, 130 times larger than the rest-frame *V*-band effective radius (Figure 3). This extended low velocity CO emission could also be associated with feedback processes (for example, previously ejected gas), but we cannot rule out the hypothesis that it represents molecular gas ejected from discs during previous stages of the merger.

We assume that the unresolved CO component is associated with dense gas still actively forming stars. The mass of this active component is estimated as $M_{\text{H}_2} = (3.1 \pm 0.6) \times 10^9\,M_\odot$ assuming thermalised CO $J = 2 \to 1$ emission and $\alpha = 0.8\,M_\odot\,(\text{K}\,\text{km}\,\text{s}^{-1}\,\text{pc}^2)^{-1}$, where $\alpha L'_{\text{CO}} = M_{\text{H}_2}$, appropriate for the conditions in the nuclear regions of ultraluminous infrared galaxies[16]. The infrared-to-CO luminosity $L_{\text{IR}}/L'_{\text{CO}} \approx 800\,L_\odot\,(\text{K}\,\text{km}\,\text{s}^{-1}\,\text{pc}^2)^{-1}$ is close to the upper limit predicted for radiation pressure limited star formation[10]. The masses in the extended low velocity ($|\Delta V| < 200\,\text{km}\,\text{s}^{-1}$) and high velocity ($\Delta V = 200\text{--}1000\,\text{km}\,\text{s}^{-1}$) wing components are $M_{\text{H}_2} = (1.1 \pm 0.5) \times 10^9\,M_\odot$ and $M_{\text{H}_2} = (0.6 \pm 0.2) \times 10^9\,M_\odot$ respectively. Combined, the extended CO emission represents approximately 35 per cent of the total gas mass. An uncertainty here is the choice of $\alpha$; we assume a conservative value of $\alpha = 0.34\,M_\odot\,(\text{K}\,\text{km}\,\text{s}^{-1}\,\text{pc}^2)^{-1}$ for the extended components, which assumes local thermodynamic equilibrium and is applicable to the optically thin case of turbulent gas associated with a wind, assuming an CO/$\text{H}_2$ abundance of $10^{-4}$ and typical excitation temperature of $30\,\text{K}$[17].

If the extended CO-emitting gas forms a foreground screen, then the hydrogen column density can be inferred in the limit where the column is dominated by molecular gas. As with the molecular mass estimate, one must adopt a value for the 'X-factor' which relates CO emission to column density $X_{\text{CO}} = N_{\text{H}_2}/W_{\text{CO}}$, with $N_{\text{H}_2}$ in units of $\text{cm}^{-2}$ and $W_{\text{CO}}$ in units of $\text{K}\,\text{km}\,\text{s}^{-1}$. We assume $X_{\text{CO}} = 1.6 \times 10^{19}\,\text{cm}^{-2}\,(\text{K}\,\text{km}\,\text{s}^{-1})^{-1}$, for the same assumptions as $\alpha$ for the extended components described above, which yields $N_{\text{H}} \approx 2 \times 10^{20}\,\text{cm}^{-2}$. Independently, the extinction of the stellar and nebular emission can be estimated from the relative intensities of the Balmer lines, which indicate an extinction of $A_V \approx 0.5\,\text{mag}$ to the young, compact stellar population (assuming Milky Way abundances[18]). This corresponds to a column density of $N_{\text{H}} \approx 10^{21}\,\text{cm}^{-2}$, in reasonable agreement with the value estimated from the cold gas. A plausible scenario is that an outflow, or series of outflows, launched from the nuclear starburst has purged the ISM of the stellar bulge, sweeping cold gas and dust into the halo. This 'blow out' phase will rapidly truncate star formation in the bulge on a timescale comparable to the dynamical time, exposing a bright shell of young stars around the nuclear starburst, which will consume the remaining molecular gas within $10\,\text{Myr}$ given the gas supply and consumption rate.

For cold gas to be driven to large galactocentric distance, the wind driving mechanism must be favourable to the survival of cold clouds[9]. Originally it was thought that cold material would be ejected along with the hot gas associated with the explosions of supernovae[19], but cold clouds entrained in such outflows are predicted to be destroyed on timescales of Myr and quickly incorporated into the hot flow[20]. Alternatively, stellar radiation pressure on dust grains can accelerate cold gas several Myr before the first cluster supernovae explode, without subjecting it to the deleterious effects of a hot conductive atmosphere[9]. This mechanism has already been shown to be the most likely driver in local starbursts exhibiting sub-kpc molecular outflows[2]. After exiting the galaxy, the cold gas interacts with the potentially hot halo atmosphere. Although the hydrodynamic interaction between a hot ($10^6$ K) atmosphere and high velocity cold–cool ($10$–$10^4$ K) clouds is complex[21], these observations suggest that ejected molecular gas can survive in such an environment for timescales of at least approximately 10 Myr.

If the maximum velocity ($\sim 1000\,\mathrm{km\,s^{-1}}$) in the redshifted wing is representative of the deprojected velocity of the outflow[22] then the extended CO emitting gas carries kinetic power $P_\mathrm{K} = (2.6 \pm 0.8) \times 10^{36}$ W. The mass outflow rate is $\dot{M}_{\mathrm{H}_2} = 80 \pm 25\,M_\odot\,\mathrm{yr}^{-1}$, implying a 'mass-loading' factor $\dot{M}_{\mathrm{H}_2}/\mathrm{SFR} \approx 30\%$. In a momentum driven wind, the mass loading factor scales inversely with outflow velocity[23], and our results are roughly consistent with an extrapolation of the linear correlation between $\dot{M}_{\mathrm{H}_2}/\mathrm{SFR}$ and $v$ found for local pure starbursts[2]. The rate of momentum input available from stellar radiation pressure in the single scattering limit is $L/c \approx 4 \times 10^{30}$ N[24], but note that this will be larger in the case where the medium is optically thick to far-infrared photons, scaling with the optical depth $\tau_\mathrm{FIR}$[9]. The outflow momentum flux is $v\dot{M}_{\mathrm{H}_2} = (4.8 \pm 1.9) \times 10^{30}$ N implying that the energy of the outflow is compatible with the budget available from star formation alone. Material travelling at $1000\,\mathrm{km\,s^{-1}}$ takes less than 10 Myr to reach the measured $r = 8 \pm 2$ kpc extent of the outflow. Again, this estimate is uncertain due to projection effects, but the timescale is in broad agreement with the age of the young stellar population. Fitting of the rest-frame ultraviolet/optical spectra reveals that 90% of the stellar luminosity is contributed by a population of age 6 Myr or younger. We cannot rule out the possibility that the outflow was launched by an AGN that has since 'switched off', but the current observations indicate that the outflow is compatible with pure stellar feedback.

A key goal in galaxy evolution studies has been to understand the coupling between various forms of energy and momentum injection and the cold ISM, as well as their relative efficacies as feedback channels. The molecular outflow in SDSS J0905+57 at $z = 0.7$ is extended on a much larger scale than has been previously observed in local starburst galaxies exhibiting molecular winds[1,2]. In part, this could be related to the ultra-compact morphology and extreme nature of the system, which has a star formation rate density orders of magnitude larger than the local systems. In local galaxies, where small-scale radiatively driven molecular winds have been observed, only a few per cent of the total gas reservoir is involved in the outflow[1,2], whereas in SDSS J0905+57 up to a third of the total molecular gas reservoir appears to have been ejected. These observations are evidence that pure stellar feedback can affect the evolution of a galaxy as a whole on short (Myr) timescales by directly removing the dense material required for star formation, and is therefore competitive with AGN feedback[25–27] as a mechanism to regulate stellar mass growth and redistribute baryons in massive galaxies.

## Acknowledgements

J.E.G. acknowledges support from the Royal Society through a University Research Fellowship. A.M.D. acknowledges support from The Grainger Foundation. G.H.R. acknowledges the support of the Alexander von Humboldt Foundation and the hospitality of the Max Planck Institute for Astronomy. A.L.C. acknowledges funding from NSF CAREER grant AST-1055081. The authors thank Elias Brinks, Norman Murray, Desika Narayanan, Roberto Neri and Fabian Walter for useful advice and discussions. This work is based on observations carried out with the IRAM Plateau de Bure Interferometer. IRAM is supported by INSU/CNRS (France), MPG (Germany) and IGN (Spain). Some of the data presented herein were obtained at the W. M. Keck Observatory, which is operated as a scientific partnership among the California Institute of Technology, the University of California and the National Aeronautics and Space Administration. The Observatory was made possible by the generous financial support of the W. M. Keck Foundation.

## Author contributions

J.E.G. and R.C.H. led the original IRAM observation proposals. All authors assisted with the data analysis and writing of the manuscript. J.E.G. and M.K. led the IRAM data reduction and analysis, J.M., A.M.D. and C.A.T. led the analysis of the Keck spectroscopy and P.H.S. led the morphological analysis of the Hubble Space Telescope imaging.

## Author information

Reprints and permissions information is available at www.nature.com/reprints. The authors declare no competing financial interests. Correspondence and requests for materials should be addressed to J.E.G. (j.geach@herts.ac.uk).

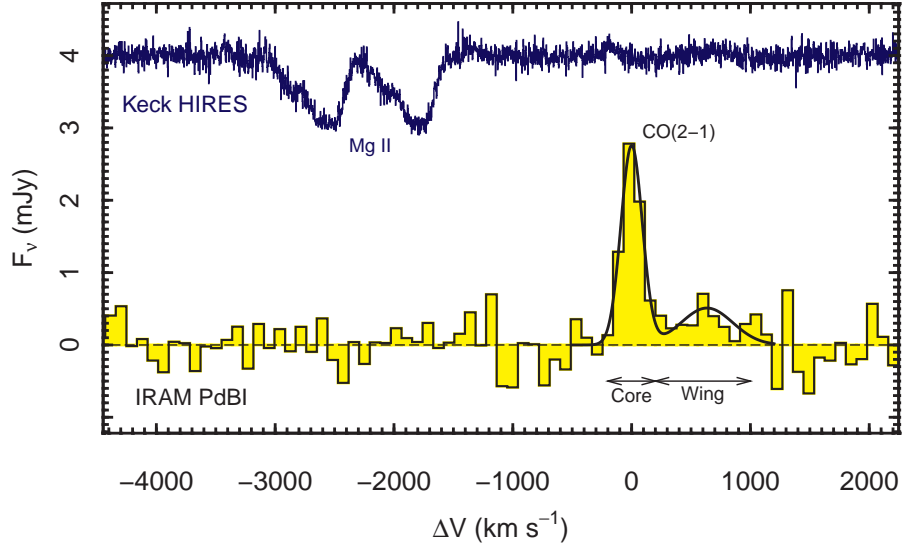

Figure 1: 2 mm spectrum of SDSS J0905+57 obtained with IRAM PdBI. The CO(2–1) line is fit with a gaussian profile of width $200\,{\rm km\,s^{-1}}$ (full width at half maximum) and peaks at a redshift consistent with the stellar absorption lines. There is evidence for significant CO emission in a high velocity wing that extends up to $1000\,{\rm km\,s^{-1}}$ from the core line, which could indicate a high velocity outflowing molecular gas component; we jointly fit this with a gaussian profile. SDSS J0905+57 is also driving a high-velocity outflow of ionised gas, as revealed by strongly blue-shifted Mg II $\lambda 2796$, $\lambda 2803$Å doublet absorption observed in the Keck HIRES rest-frame ultraviolet spectrum, shown here on the same velocity (but arbitrary flux) scale, relative to the 2796Å line.

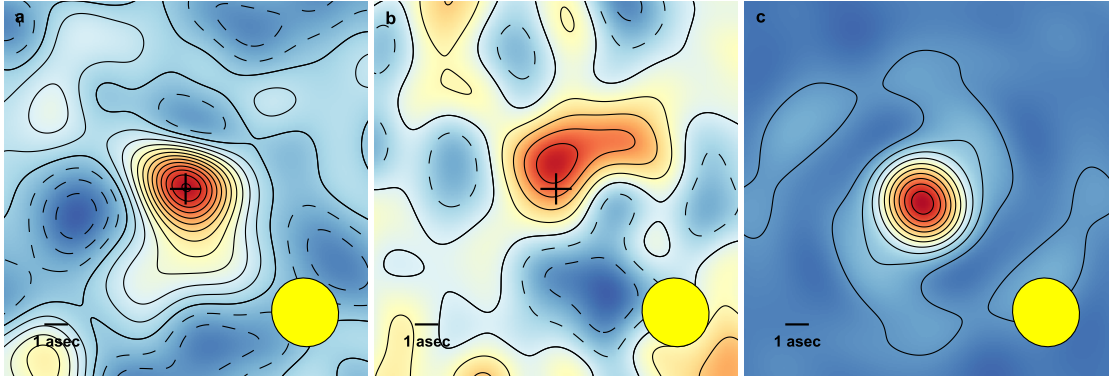

Figure 2: Maps of carbon monoxide emission. The panels show cleaned CO(2–1) line maps averaged over (a) the full width at zero intensity core line ($\pm 200\,{\rm km\,s^{-1}}$) and (b) the wing emission over 200–1000 $\,{\rm km\,s^{-1}}$. Contours are spaced at multiples of the root mean squared noise level in the maps, with dashed contours tracing negative deviations. The wing emission (b) peaks $(1.2 \pm 0.3)''$ or $8 \pm 2$ kpc from the core line with a peak flux of $0.43 \pm 0.09$ mJy ($4.8\sigma$). (c) shows the shape of the dirty beam, with contours at 10–90% of peak. The yellow ellipse in all panels represents the FWHM of the dirty beam, approximately $3''$.

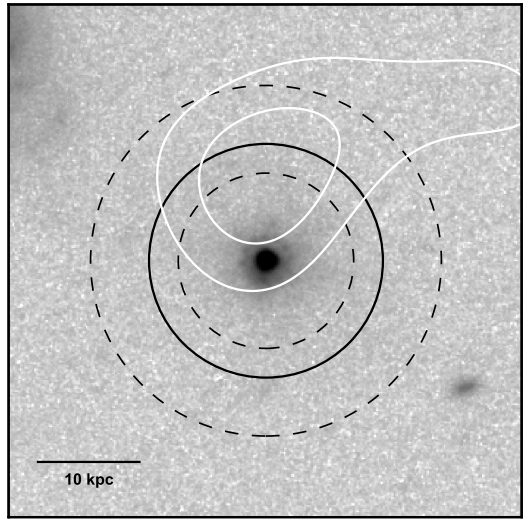

Figure 3: Optical image of SDSS J0905+57 from the Hubble Space Telescope. The optical image reveals the rest-frame *V*-band morphology of the target obtained with the Wide Field Camera 3 Ultraviolet Imaging Spectrograph (F814W filter). The galaxy has an effective radius of 100 pc for the rest-frame *V*-band light. White contours show the 3–4$\sigma$ levels for CO emission averaged over 200–1000 km s$^{-1}$. The black circle shows the CO half-power size when averaged over the FWZI of the core line. Dashed lines show the 1$\sigma$ confidence bounds for the half power size.

# Methods

Linked to the online version of the paper at www.nature.com/nature.

## Target information

The target SDSS J0905+57 at $09^h\, 05^m\, 23.7^s$, $+57°\, 59'\, 12.6''$ ($z = 0.712$) is a galaxy with the fastest Mg II outflow velocity of a larger sample of similar galaxies originally selected post-starburst spectral features[8,28]. The total stellar mass, estimated from fitting of the rest-frame ultraviolet-to-near-infrared spectral energy distribution (SED), is $M_\star = 5.5 \times 10^{10} M_\odot$, and analysis of the rest-frame near-ultraviolet/optical spectra show that 90% of the stellar emission is contributed by populations younger than 6 Myr, representing 20% of the stellar mass. The rest-frame $V$-band size in the Hubble Space Telescope imaging was measured following the methods presented elsewhere[15], which we briefly describe here. To parametrise the compactness of the galaxy, we fit Sérsic and Sérsic + PSF models to the galaxy, with the Sérsic index frozen to $n = 4$, and its nearby masked field (a $28'' \times 28''$ box centered on the galaxy) with GALFIT version 3[29]. The effective radius is $r_e = 94$ pc, with two thirds of the rest-frame ultraviolet/optical light unresolved by the Hubble Space Telescope.

The bolometric luminosity is estimated through extrapolation of the mid-infrared (Wide-Field Infrared Survey Explorer 12$\mu$m, 22$\mu$m) photometry[8], using appropriate spectral energy distribution templates that provide an estimate of the far-infrared emission assuming these compact star-forming galaxies conform to the typical cool dust emission seen in star-forming galaxies[30]. The star formation rate derived from the total (integrated 8–1000 $\mu$m) dust emission is in agreement with that derived from the fitting of the rest-frame 0.1–3$\mu$m SED, with proper treatment of differential dust obscuration of the stellar emission[8,31]. The measured $L_{\rm IR} \approx 1.2 \times 10^{39}$ W corresponds to a star formation rate of 260 $M_\odot\,{\rm yr}^{-1}$ assuming a Chabrier initial mass function[32] (a Salpeter initial mass function would increase the SFR by 80%). If 50% of the star formation occurs within the effective radius measured above, then the projected star formation rate density is $\Sigma_{\rm SFR} \approx 4700\, M_\odot\,{\rm yr}^{-1}\,{\rm kpc}^{-2}$. As a guide to the level of uncertainty in derived properties, which depend largely on template fitting, both the stellar mass and infrared luminosity are estimated to be accurate to within a factor of two.

## IRAM observations and data reduction

SDSS J0905+57 was observed on 6–12 May 2013 and 10 December 2013 as part of IRAM PdBI projects W09A and X09C. PdBI was in compact (D) configuration with baselines of 25–140 m. We used the WideX correlator, targeting the redshifted CO(2–1) line at $\nu_{\rm obs} \approx 134$ GHz in the 2 mm band, recording dual polarization. The mean system temperature was $T_{\rm sys} = 80$–120 K and precipitable water vapour was in the range 2–6 mm. The sources 3C84, 3C279, 3C454.3 2200+420 and 0851+202 were used for bandpass calibration, and sources 0954+658 and 0917+624 were used for phase/amplitude calibration. We rejected scans for which the phase r.m.s. deviated more than 45 degrees from the calibration solution. Finally, the source MWC349 was used for flux calibration (accuracy 5–10% at 2 mm). The final r.m.s. noise in 20 MHz (45 km s$^{-1}$) channels is $\sigma = 0.4$ mJy. The package GILDAS[33] was used for data calibration, mapping and analysis.

## Line detection

We first map the $uv$ visibilities into the image plane to create a spectral cube from which we extract a spectrum from a single $0.6''$ pixel at the phase tracking centre. This spectrum is shown in Figure 1, with the CO(2–1) emission line strongly detected at the expected frequency. The line is well modelled by a single gaussian, with peak $S = 2.8 \pm 0.3$ mJy and $\sigma_{\rm FWHM} = 200 \pm 35$ km s$^{-1}$. Uncertainties on the profile fit parameters such as line centres and widths are estimated in the following way: first, we estimate the uncertainty per channel by extracting spectra from 100 random locations close to the phase tracking centre, but in line-free parts of the data cube, and then evaluate the r.m.s. variation in the signal for each channel. Under the assumption that the noise is drawn from a gaussian distribution, $\mathcal{G}$ with mean of zero and a $1\sigma$ scale equivalent to the r.m.s., we generate 1000 realisations of the target spectrum, each time adding flux drawn randomly from $\mathcal{G}$ to each channel. The integrated flux and line fits are re-evaluated for each of the 1000 realisations, and the standard deviation of the derived values are taken to be the $1\sigma$ uncertainties on the fit parameters. The integrated line flux over the 200–1000 km s$^{-1}$ wing is $S\Delta V = 0.26 \pm 0.07$ Jy km s$^{-1}$, corresponding to $L'_{\rm CO} = (1.8 \pm 0.5) \times 10^9$ K km s$^{-1}$ pc$^2$. When integrated over the core line, $|\Delta V| < 200$ km s$^{-1}$, the integrated line flux in the extended and unresolved components (see below) is $S\Delta V = 0.46 \pm 0.20$ Jy km s$^{-1}$ and $S\Delta V = 0.56 \pm 0.12$ Jy km s$^{-1}$ respectively, corresponding to $L'_{\rm CO} = (3.1 \pm 1.3) \times 10^9$ K km s$^{-1}$ pc$^2$ and $L'_{\rm CO} = (3.8 \pm 0.8) \times 10^9$ K km s$^{-1}$ pc$^2$.

**Red, spatially offset high-velocity wing emission**

The spectrum shown in Figure 1 reveals evidence of CO emission red-ward of the core gaussian line in a broad wing that extends to approximately 1000 km s$^{-1}$. Averaging the 1D spectrum shown in Figure 1 over 200–1000 km s$^{-1}$ yields an average flux density of $S = 0.34 \pm 0.09$ mJy for this feature. When the spectral cube is collapsed over the same channels, the peak of this wing emission is spatially offset $(1.2 \pm 0.3)''$ from the peak of the core line, with an average flux density of $S = 0.43 \pm 0.09$ mJy (Figure 2). The positional uncertainty is estimated as $\sigma = 0.3 \times {\rm FWHM}/{\rm SNR}$[34,35], which is consistent with the positional error derived from a simulation involving the input of 1000 model sources of the same flux level into the noise map. The scatter (standard deviation) in recovered positions is $0.3''$ in R.A. and Dec. with no systematic positional offset.

An alternative way of assessing the significance of this feature is to use the random noise realisations of the data cube as described above, but excluding the source; i.e. just considering the noise component. This takes into account the possibility that the noise in consecutive channels might be correlated. We then evaluate the rate at which we measure flux densities of $S \geq 0.43$ mJy averaged over the same channels as the observed wing feature. Using $10^7$ noise realisations, we find a rate of $1.4 \times 10^{-6}$, consistent with the $4.8\sigma$ significance of the detection (we find the same rate of negative fluctuations of equivalent magnitude).

**Resolved core line emission**

When averaged over the core gaussian line, the map suggests that the CO emission is extended compared to the beam (Figure 2). To verify this, and to evaluate the size of the emitting region, we examine the velocity averaged signal amplitude as a function of baseline separation (i.e. synthetic aperture size) in the *uv* plane, since an unresolved source will have a flat amplitude-radius profile. We average over the full line width, $|\Delta V| < {\rm FWHM}$, approximately corresponding to the full width at zero intensity (400 km s$^{-1}$). Figure ED1 shows that the flux distribution, evaluated as the average amplitude in radial bins of width 50 m, deviates from a flat distribution at baseline separations shorter than approximately 100 m. We see evidence for this in both independent observations of the target. As described above, we have imposed strict flagging throughout, rejecting scans for which the phase r.m.s. deviated more than 45 degrees from the calibration solution, and therefore any smearing of the signal related to phase calibration errors will occur on scales of approximately $1.5''$ (roughly half the size of the synthesised beam). This is a strong indication that the extended emission is real and not a result of seeing, however as an additional check, we examine the profiles of the phase calibrators themselves.

If the extended emission is due to a phase calibration issue, in which point source emission is artificially smeared out, then we would expect to see a similar extended profile around the calibrators as well as the target. The first test we perform is to measure the average amplitude as a function of *uv* radius for the main calibrator common to both projects, source 0917+624. Figures ED2 and ED3 summarise the results with the map and amplitude-radius profile of the main calibrator, with the master phase calibration solution applied, indicating consistency with an unresolved source. During project X09C, two calibrators were observed: 0917+624 and 0954+658. This gives us the opportunity to derive a phase calibration solution excluding one of these sources (0954+658) and then apply that solution to the excluded source. This is a robust test, since 0954+658 has not contributed to the phase solution, and can be treated as an independent source. We show the map for 0917+624 and 0954+658 (with the latter 'blindly' treated with the phase solution derived from the main calibrator) in Figure ED2, and the corresponding amplitude-radius profiles in Figure ED3. Again, the results indicate unresolved sources, which would not be the case if phase calibration errors are responsible for the extended emission observed in the target source. These results give us further confidence that the extended emission we observe in SDSS J0905+57 is real.

It is common practice to model partially resolved emission with a gaussian profile, the half power size of which can be used to characterise the size of the emitting region. We fit the flux distribution with a combination of gaussian profile and point source:

$$S(r) = S_0 \exp\left(-\frac{(\pi r b)^2}{\ln(2)}\right) + S_1 \qquad (1)$$

where $r = (u^2 + v^2)^{1/2}$, $S_0$ is the peak amplitude of the extended component, $b$ is the half width half power size and $S_1$ is the amplitude of the unresolved component. To improve signal-to-noise, we were required to average the visibilities into three bins, so to reduce the number of free parameters in the model, we make the following assumptions: first, we assume that the flux density on the longest baselines is representative of the flux density of the unresolved component, $S_1 = 1.6 \pm 0.3$ mJy. We then fix $S_0 = (2.8 - S_1)$ mJy, corresponding to the peak emission measured in the spectrum described above, corrected for a contribution from an unresolved component. With these parameters fixed, we then perform a $\chi^2$ minimisation between the model and the data to find the best-fitting half power size, $b$. In order to find the range of acceptable values for the fit, taking into account the uncertainties on the assumptions for $S_0$ and $S_1$, we perform a Monte Carlo simulation, re-fitting the data by sampling $S_0$ and $S_1$ from gaussian distributions with widths concordant with the $1\sigma$ errors on the flux densities. We perform 1000 trials and take the mean and standard deviation in best-fit for each trial $b$ as the final size estimate. We find $b = 62 \pm 21$ metres or $\theta = 1.6^{+0.8}_{-0.4}$ arcseconds for $\lambda = 2.23$ mm, with a best fit $\chi^2 = 0.07$ for 1 degree of freedom, implying that the data are over-fit. Additional observations that would allow us to increase the number of bins in the amplitude–radius plane would improve our constraints on the size of the extended emission. Nevertheless, the $\chi^2$ for the null hypothesis (that the source is unresolved) is $\chi^2 = 20.8$, with the $\Delta\chi^2$ corresponding to a significance of $4.7\sigma$. Thus, we can rule out the point source only model with reasonable confidence.

**Feedback energetics**

SDSS J0905+57 is forming stars close to the theoretical upper limit for radiation pressure (Eddington limited) star formation, with $L_{\rm IR} \approx L_{\rm Edd} \propto GcL'_{\rm CO}/\kappa$ for optically thick dust emission ($\tau_{100\mu m} > 1$), where $\kappa$ is the Rosseland-mean dust opacity ($\kappa \approx 1000 f_{\rm dg}$ cm$^2$ g$^{-1}$, where $f_{\rm dg}$ is the dust-to-gas mass ratio, typically $\frac{1}{50}$–$\frac{1}{150}$[10]). The high star formation rate density of SDSS J0905+57 is above the threshold necessary for launching a radiation pressure driven wind[8–10,36].

The clear evidence that the galaxy has a high velocity outflow, traced by Ca II, Fe II and Mg II (Diamond-Stanic et al. in prep) is an unambiguous signature of an energetic, gaseous outflow launched in the recent past. The observations presented here imply that molecular gas has also been ejected by feedback processes. In the following we estimate the rate of momentum input from the starburst and compare this to the energetics of the wind in order to determine if radiation pressure is a viable power source for the outflow. Considering gas associated with the high velocity wing, we determine the mass outflow rate as follows:

$$\dot{M}_{H_2} = \frac{M_{H_2} v}{r} \quad (2)$$

where $M_{H_2}$ is the gas mass in the outflow, $v$ is the outflow velocity, and $r$ is the radius of the outflow. We make the conservative assumption that the gas is travelling at $1000\,\mathrm{km\,s^{-1}}$, assuming that the maximum velocity extent of the wing is representative of the deprojected velocity of the outflow[22]. The mass outflow rate is $\dot{M}_{H_2} = 80 \pm 25\,M_\odot\,\mathrm{yr}^{-1}$ and it follows that the kinetic power in the outflow is

$$P_K = \frac{\dot{M}_{H_2} v^2}{2} \quad (3)$$

and yields $P_K = (2.6 \pm 0.8) \times 10^{36}\,\mathrm{W}$, several orders of magnitude lower than the bolometric luminosity. Again, it is important to highlight that the error bars do not reflect the systematic uncertainty from the choice of $\alpha$ that we use to estimate $M_{H_2}$ from $L'_{CO}$. In these calculations we have assumed $\alpha = 0.34\,M_\odot\,(\mathrm{K\,km\,s^{-1}\,pc^2})^{-1}$ that could be appropriate for the optically thin conditions in a turbulent molecular outflow[17]. The conversion factor $\alpha$ is defined as $M_{H_2}/L'_{CO}$ for $J = 1 \to 0$, thus a correction is required for higher-order transitions (as are often measured in high-redshift galaxies) to account for the shape of the spectral line energy distribution (SLED) that describes the excitation of the gas. We have no constraints on the excitation state of the molecular gas in this galaxy (a wider range of transitions will be needed to constrain the SLED), and so for all components we have assumed thermalised CO emission such that $r_{21} = L'_{CO(2-1)}/L'_{CO(1-0)} = 1$.

We argue that the outflow can be driven by radiation pressure from the compact starburst, so it is critical to assess the momentum injection available to drive cold gas. In the single scattering limit a starburst's radiation pressure scales with the bolometric luminosity, $L/c$[23,24]. For SDSS J0905+57 we find $\dot{p}_{\rm rad} \approx 4 \times 10^{30}\,\mathrm{N}$. The momentum flux in the wind is $v\dot{M}_{H_2} \approx (4.8 \pm 1.9) \times 10^{30}\,\mathrm{N}$, implying that the starburst could be driving the outflow through stellar radiation pressure alone, even in the single scattering limit. Given that we have neglected other driving forces such as supernovae ram pressure, which could provide an additional factor of $\sim 3$ times the momentum input from radiation pressure[24] we conclude that it is plausible that a central compact starburst of this magnitude could drive the molecular outflow we observe.

How does the energy of the molecular outflow compare to the high-velocity outflow of warmer ionized gas traced by Mg II? There are several uncertainties in deriving an energy with the Mg II outflow, the most serious of which is the lack of any useful constraints on the geometry and scale of the outflow (although the saturated Mg II lines suggest a high covering factor possibly consistent with a shell geometry). Other uncertainties include the use of saturated Mg II lines, uncertain Mg$^+$/Mg ionisation and dust depletion corrections and Mg/H abundance. Nevertheless, for a shell of radius $r$ the mass scales as

$$M \approx 5.5 \times 10^8 M_\odot \left(\frac{r}{5\,\mathrm{kpc}}\right)^2 \left(\frac{N_H}{2 \times 10^{20}\,\mathrm{cm}^{-2}}\right)$$

and the kinetic energy

$$E \approx 3.4 \times 10^{51}\,\mathrm{J} \left(\frac{r}{5\,\mathrm{kpc}}\right)^2 \left(\frac{N_H}{2 \times 10^{20}\,\mathrm{cm}^{-2}}\right) \left(\frac{v}{2500\,\mathrm{km\,s^{-1}}}\right)^2.$$

Thus, for a shell of radius 5 kpc and column of $N_H = 2 \times 10^{20}\,\mathrm{cm}^{-2}$ there is comparable mass and approximately six times more energy in the ionized outflow compared to the molecular outflow, indicating that a contribution from supernovae is required to drive the ionised wind. A possible scenario is that the blow out phase of the final-stage merger involves two key stages: radiative feedback that first drives cold gas out of the bulge and into the halo, clearing low-density 'escape paths' for warmer gas that is then driven out by supernova detonations, several Myr later[9].

What of the other extended, low-velocity CO component associated with the core line? It is difficult to infer feedback energetics associated with the gas that is extended on scales of 12 kpc. This gas is moving with lower velocity than the outflow (i.e. similar to the circular velocity, $(GM/R)^{0.5} \approx 150\,\mathrm{km\,s^{-1}}$ with $M(r < R) = 6 \times 10^{10}\,M_\odot$ dominated by baryons and $R = 12\,\mathrm{kpc}$). If this gas was ejected in a shell that has been radiatively driven into the halo in a previous outflow event, it is likely to have fragmented and decelerated; at late times the cold clouds are subject to lower radiation and ram-pressure[9]. In this case an instantaneous measure of mass outflow rate and its connotations loses the intended meaning, and the previously expelled cold gas merely represents the time-integrated feedback history.

**Arguments for starburst feedback and against AGN feedback**

The arguments for star formation as opposed to AGN feedback for this galaxy and others like it are discussed in previous work[8,15]. For this source, we have shown that the compact starburst can produce the observed outflows, but cannot conclusively rule out that they were launched by an AGN that has since 'switched off'. However, the larger population of compact starbursts shows ubiquitous outflows with no correlation with AGN activity, pointing to star formation as the most likely driver for the observed feedback in this and similar systems. Furthermore, there are two main pieces of observational evidence that implies that SDSS J0905+57 does not contain an energetically dominant AGN:

1. AGN activity is commonly diagnosed via high excitation emission lines, especially when X-ray observations are unavailable (as is the case for SDSS J0905+57) as it is typically assumed that the AGN is the only possible source for considerable numbers of high-energy photons. We have observational constraints on the emission lines [O III]$\lambda$5007Å, [Ne III]$\lambda$3869Å and [Ne V]$\lambda$3426Å. We measure $\log\left([\mathrm{OIII}]/\mathrm{H}\beta\right) = 0.19$ and $L_{[\mathrm{OIII}]} = 7 \times 10^{34}$ W. No [Nev]$\lambda$3426Å is detected ($L_{[\mathrm{NeV}]} < 2.25 \times 10^{33}$ W, $1\sigma$ upper limit), and the [Ne III] line is weak, $L_{[\mathrm{NeIII}]} = 1.64 \times 10^{33}$ W ($3\sigma$ detection). Very compact starburst galaxies can produce slightly elevated excitation emission lines with these observed properties[37] and these observations are consistent with a starburst dominated system at $z \approx 0.7$.

2. Mid-infrared observations can be used to assess obscured AGN activity. The 3.6–4.5$\mu$m colour from Spitzer observations of SDSS J0905+57 is $[3.6] - [4.5] = 0.48$ (Vega magnitudes), which is in a transition region between star-forming galaxies and AGN[38,39]. Note that the 5.8–8.0$\mu$m colour was unavailable in the Spitzer Warm Mission observations. We can also estimate $L_{\mathrm{bol}}$ from $L_{[\mathrm{OIII}]}$, in the limit where all the [OIII] is produced from an AGN. Keeping in mind that $L_{[\mathrm{OIII}]}$ is likely heavily contaminated by star formation, using $L_{\mathrm{bol}}/L_{[\mathrm{OIII}]} = 600$[40], we find $L_{\mathrm{bol}} = 4 \times 10^{37}$ W, approximately 4 per cent of the total infrared luminosity.

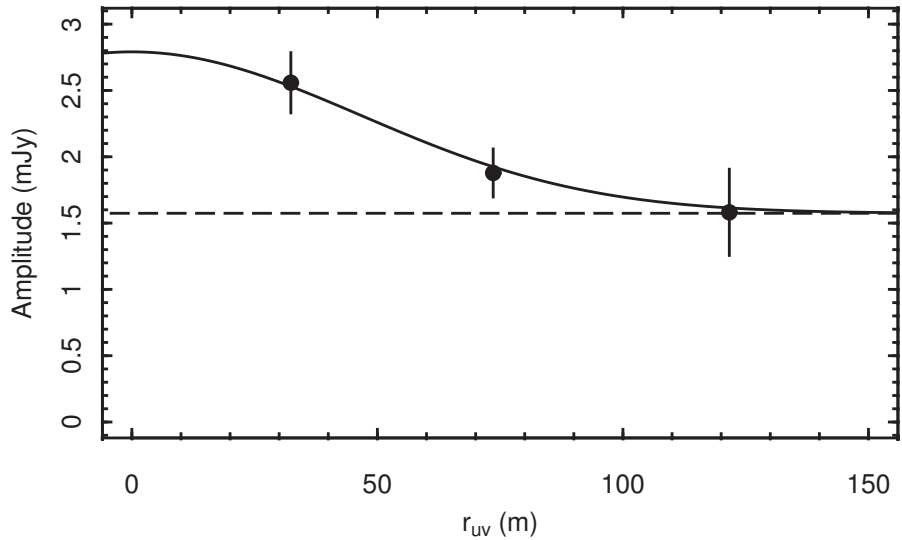

Figure ED1: Average signal amplitude as a function of baseline separation for CO emission over the core line, $\Delta V = \pm 200\,\text{km}\,\text{s}^{-1}$. This reveals a significant deviation from a flat profile indicating that the CO emission is partially resolved. The profile is accurately modelled by a combination of a point source (a delta function in the $uv$ plane) and a gaussian profile with half power radius of $\sim 2''$. A point source-only model can be ruled out at the $4.7\sigma$ level. Error bars show the $1\sigma$ confidence range, and derived as $w^{-0.5}$ where $w$ is the weight $w = \Delta\nu t / T_{\text{sys}}^{\text{ab}}$, where $\Delta\nu$ is the channel width, $t$ is the integration time and $T_{\text{sys}}^{\text{ab}}$ is the product of the system temperature of two antennas.

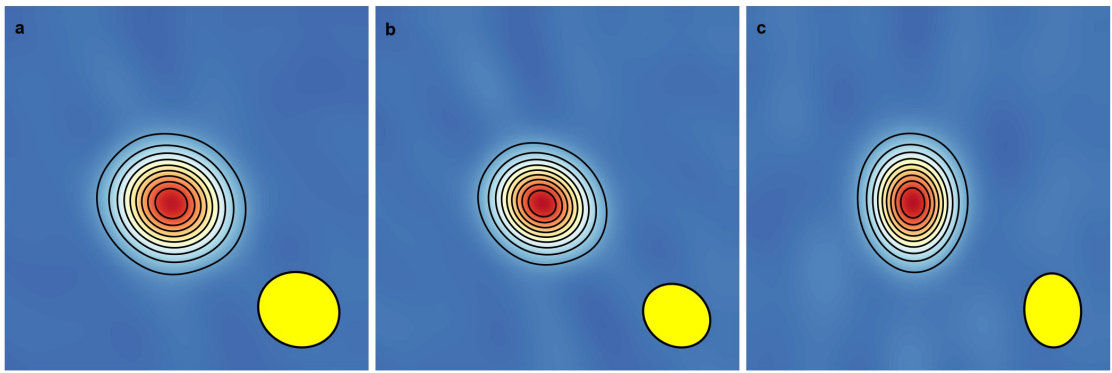

Figure ED2: Clean maps of the phase calibrators 0917+624 and 0954+658. The full width half maximum beam shape indicated as an ellipse. Contours are at levels of 10–90% of peak. (a) phases have been calibrated with a solution derived from calibrators observed over both projects, (b) the same phase solution has been applied to observations of 0917+624 observed in project X09C only, (c) the calibrator 0954+658 has been calibrated with a solution derived from 0917+624 observed during X09C only. All sources have a profile that is matched to the synthesised beam, with no evidence of extended emission.

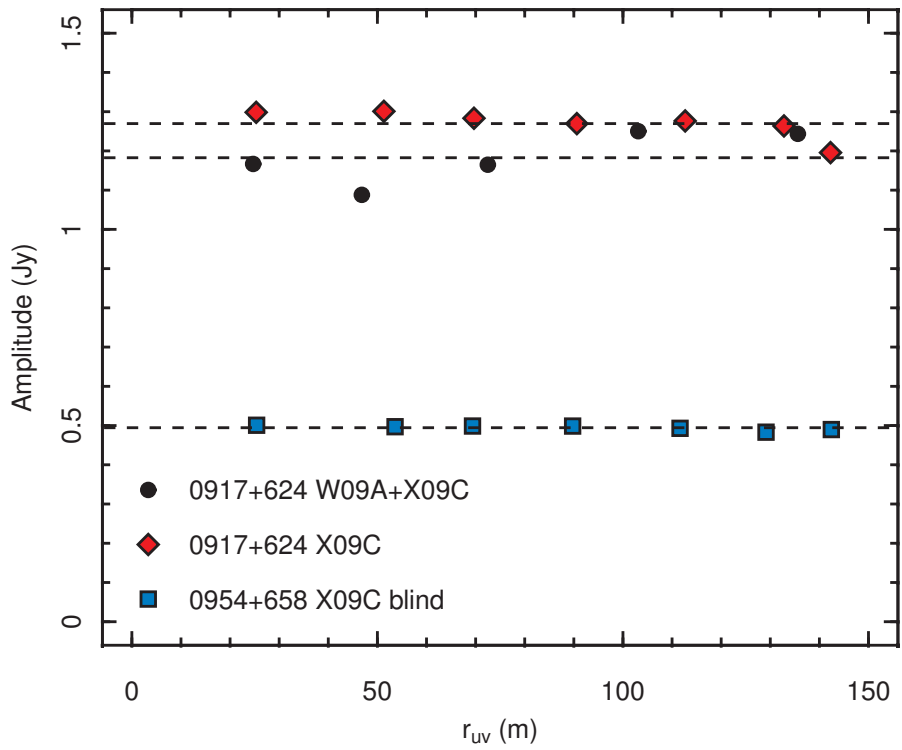

Figure ED3: Circularly averaged amplitude-radius profiles of the maps shown in Figure ED2 in the *uv* plane. All profiles are consistent with unresolved emission (flat profiles). Note that observing conditions were slightly better during project X09C than W09A.